\documentclass[%
 reprint,
superscriptaddress,
 amsmath,amssymb,
 aps,prl
]{revtex4-1}

\usepackage[utf8]{inputenc}
\usepackage{todonotes}
\usepackage{graphicx}%
\usepackage{dcolumn}%
\usepackage{bm}%
\usepackage[normalem]{ulem}
\usepackage{microtype}
\usepackage{xcolor}
\usepackage{amsmath,amssymb,amsfonts,mathtools,cancel}
\usepackage{multirow}
\usepackage{comment}
\usepackage{hyperref}

\usepackage[T1]{fontenc}
\usepackage{newtxtext}

\newcommand{\subfigimg}[3][,]{%
  \setbox1=\hbox{\includegraphics[#1]{#3}}%
  \leavevmode\rlap{\usebox1}%
  \rlap{\hspace*{0pt}\raisebox{\dimexpr\ht1-2\baselineskip}{#2}}%
  \phantom{\usebox1}%
}

\newcommand{\figLabel}[1]{\textbf{\textsf{\MakeLowercase{#1}}}}  %
\newcommand{\figLabelCapt}[1]{\textbf{\textsf{\MakeLowercase{#1}}}}  %

\newcommand{\refSub}[2]{\hyperref[#2]{\ref{#2}\figLabelCapt{#1}}}

\newcommand{\figrefsub}[2]{Fig.~\refSub{#2}{#1}}

\usepackage[paperwidth=210mm,
            paperheight=297mm,
            left=20mm,
            top=10mm,
            textwidth=170mm,
            marginparsep=3mm,
            marginparwidth=30mm,
            textheight=730pt,
            footskip=50pt]
           {geometry}

\newcommand{\kb}{k_{\rm B}}

\newcommand{\avg}[1]{\ensuremath{\langle{#1}\rangle}}
\newcommand{\abs}[1]{\ensuremath{\|{#1}\|}}
\newcommand{\pd}[2]{\ensuremath{\dfrac{\partial {#1}}{\partial {#2}}}}
\newcommand{\pds}[3]{\ensuremath{\dfrac{\partial^2 {#1}}{\partial {#2}\partial{#3}}}}

\newcommand{\ma}{\ensuremath{\mu_{A}}}
\newcommand{\mb}{\ensuremath{\mu_{B}}}
\newcommand{\na}{\ensuremath{n_{A}}}
\newcommand{\nb}{\ensuremath{n_{B}}}
\newcommand{\ca}{\ensuremath{c_{A}}}
\newcommand{\cb}{\ensuremath{c_{B}}}

\newcommand{\br}{\ensuremath{\textbf{r}}}
\newcommand{\bk}{\ensuremath{\textbf{k}}}

\usepackage[T1]{fontenc}
\usepackage{titlesec}
\titleformat{\paragraph}[runin]
{\bfseries}{\theparagraph}{1em}{}

\begin{document}

\preprint{APS/123-QED}

\title{
Computing chemical potentials of solutions from structure factors
}%

\author{Bingqing Cheng}
\email{bingqing.cheng@ist.ac.at}
\affiliation{Institute of Science and Technology Austria, Am Campus 1, 3400 Klosterneuburg, Austria}

\date{\today}%

\begin{abstract}
The chemical potential of a component in a solution is defined as the free energy change as the amount of the component changes. Computing this fundamental thermodynamic property from atomistic simulations is notoriously difficult, because of the convergence issues in free energy methods and finite size effects. This paper presents the S0 method, which can be used to obtain chemical potentials from static structure factors computed from equilibrium molecular dynamics simulations under the isothermal-isobaric ensemble. The S0 method is demonstrated on the systems of binary Lennard-Jones particles, urea--water mixtures, a NaCl aqueous solution, and a high-pressure carbon-hydrogen mixture.
\end{abstract}
\maketitle

The chemical potential of a solute as a function of its concentration 
is pivotal for understanding
many important physical and biological processes such as osmosis, the solvation of organic molecules,
the behavior of electrolytes,
precipitation of crystals from solutions,
and phase equilibria of mixtures.

Despite the massive importance, computing the chemical potentials of solutions from atomistic simulations 
is notoriously difficult
and has not become a routine calculation.
This is because that the existing ways
~\cite{sanz2007solubility,paluch2010method,lisal2005molecular,moucka2011molecular,perego2016chemical,joung2008determination,Li2017,Li2018,Vinutha2021}
have many caveats and 
are often restricted to a subset of systems.
For example, methods based on Monte Carlo particle insertion and removal ~\cite{allen2012computer,smit1989calculation} can have numerical convergence issues~\cite{perego2016chemical} and do not work for dense fluids.
Thermodynamic integration or overlapping distribution method from the real solution to a reference fluid~\cite{sanz2007solubility,paluch2010method} involves running multiple simulations with different coupling constants as well as along different thermodynamic paths,
may have singularity problems at the end points of the integration~\cite{Li2017,Li2018},
and is difficult to use for molecules with complex topology.
The direct coexistence~\cite{joung2008determination} needs long equilibration time, and only works at conditions close to the solubility limit.
Kirkwood-Buff integrals~\cite{Kirkwood1951} have severe finite size effects~\cite{CortesHuerto2016}.

Here we propose a generic, easy-to-use method (the S0 method) for computing chemical potentials of solutes at different concentrations in solutions.
This method is based on the thermodynamic relationship between 
the composition
fluctuations to derivatives of the chemical potentials to the concentration.
In practice, only the static structure factors computed from equilibrium isothermal-isobaric (NPT) molecular dynamics (MD) simulations at different solute concentrations are needed.
We benchmark the method on the systems of binary Lennard-Jones particles,
urea--water,
NaCl aqueous solution,
and a high-pressure carbon-hydrogen mixture.

\paragraph{Theory}

\begin{figure}
\includegraphics[width=0.4\textwidth]{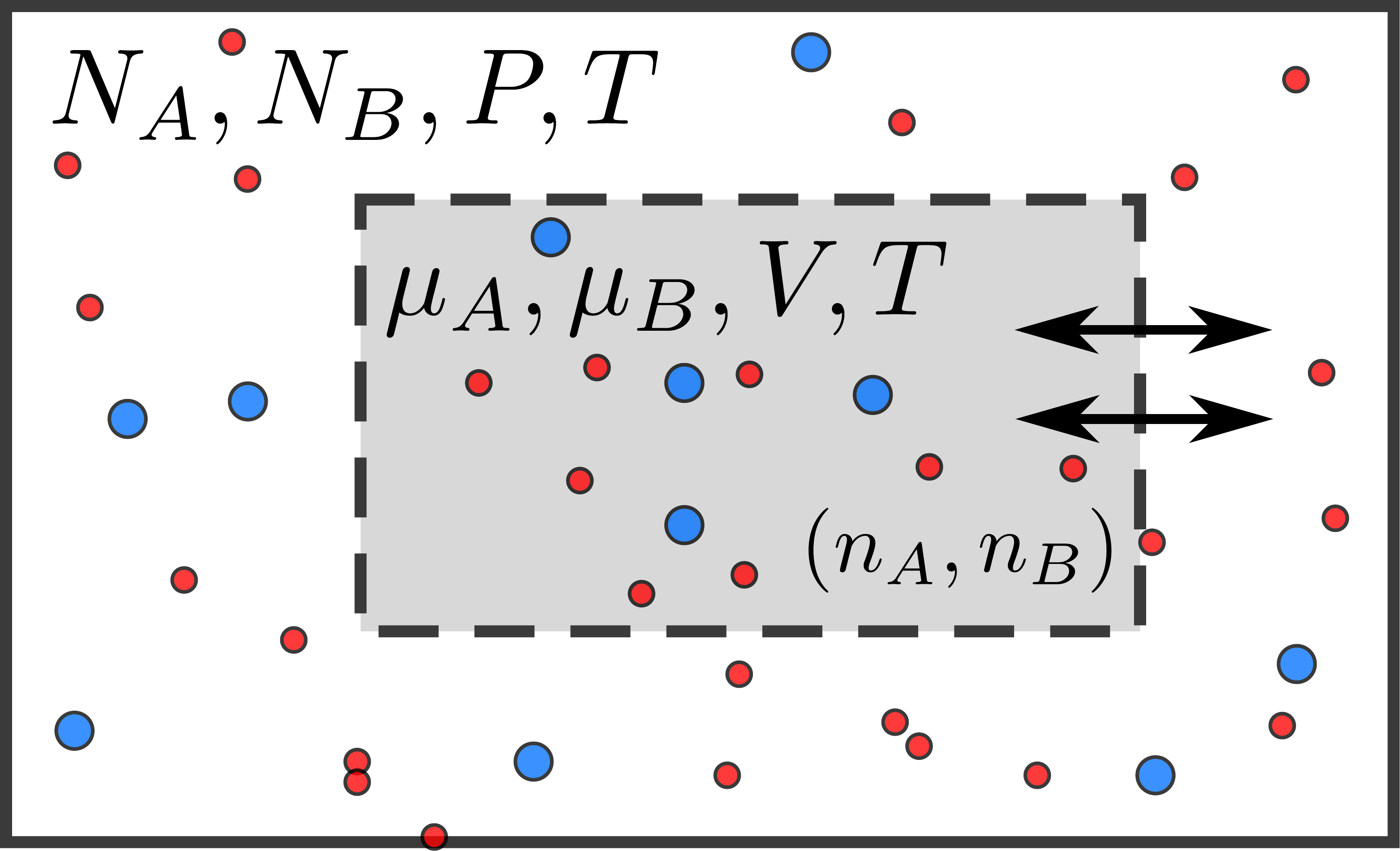}
\caption{
An illustration of the isothermal-isobaric (NPT) and the grand canonical ($\mu$VT) ensembles.
The $\mu$VT ensemble inside is only used for the theoretical derivation, and
the whole NPT ensemble is used to compute the structure factors.
}
\label{fig:ensemble}
\end{figure}

Consider a large binary particle reservoir with fixed numbers of type A and type B particles in the NPT ensemble (illustrated in Fig.~\ref{fig:ensemble}).
The Gibbs free energy for this reservoir can be expressed as
\begin{equation}
    G = \mu_A N_A + \mu_B N_B,
\end{equation}
which indicates that each particle type has a constant chemical potential.

Inside the NPT ensemble, a fixed volume $V$, which is large but much smaller than the reservoir, with permeable boundary
can be regarded a grand canonical ensemble ($\mu$VT).
The corresponding grand potential is
\begin{equation}
    \Omega= -\kb T\ln 
        \sum_{\na=0}^{N_A}  \sum_{\nb=0}^{N_B}  
            \exp {\left[\dfrac{\na \ma}{\kb T}\right]} 
    \exp {\left[\dfrac{\nb \mb}{\kb T}\right]} 
        Q(\na,\nb,V,T),
\end{equation}
where $Q$ is the canonical parition function of the $V$-region with the numbers of the two types of particles $\na$ and $\nb$.
By taking the derivatives of $\Omega$ with respect to the chemical potentials,
one gets the particle number fluctuations $\Delta \na = \na -\avg{\na}$ and $\Delta \nb = \na -\avg{\nb}$:
\begin{equation}
    \dfrac{\avg{\Delta\na\Delta\nb}}{\kb T}=  
    -\pds{\Omega}{\ma}{\mb}
    =\pd{\avg{\na}}{\mb},
    \label{eq:dndmu}
\end{equation}
where $\avg{\ldots}$ indicate the expectation value of an observable in the grand canonical ensemble.
As derived in the seminal paper of 
Kirkwood-Buff~\cite{Kirkwood1951},
such equilibrium fluctuations can be used obtain the derivatives of the $\mu$ to the at constant ($P$,$T$) conditions, e.g.
\begin{equation}
    \left(\frac {\partial \ma}{\partial c_A}\right)_{T,P}=
    \dfrac{k_BT}{c_A}
    \dfrac{1}
    {
    \avg{\Delta \na^2}/\avg{\na}
    -\avg{\Delta \na \Delta \nb}/\avg{\nb}
    },
    \label{eq:dmudc}
\end{equation}
where $\ca=\avg{\na}/V$ equals to the concentration of the type A particles in the NPT reservoir.
Ref.~\cite{Kirkwood1951} then relates the particle number fluctuations to the Kirkwood-Buff integrals of radial distribution functions.
Instead, here we take a different route and evaluate $\avg{\Delta\na\Delta\nb}$ using static structure factors detailed below.

The instantaneous density field of particle number inside the NPT ensemble is
\begin{equation}
    \rho_A(\br,t) = 
    \sum_{i_A=1}^{N_A}
    \delta(\br_{i_A}(t)-\br),
\end{equation}
where $\br_{i_A}(t)$ is the position of atom $i$ of type $A$ at time $t$,
and the average number density is
\begin{equation}
    \rho_A^{(1)}(\br) = 
    \avg{
\rho_A(\br,t)
    }_\text{NPT},
\end{equation}
where $\br_{i_A}(t)$ is the position of atom $i$ of type $A$ at time $t$,
and for isotropic systems $\rho_A^{(1)}(\br) =\ca$.
Note that $\avg{\ldots}_\text{NPT}$ is the NPT ensemble average, while $\avg{\ldots}$ without subscript indicates $\mu$VT average.
To consider the two-body correlations between the density at different points in space, $\br'$ and $\br''$:
\begin{equation}
\rho_{AB}^{(2)}(\br',\br'') = 
\avg{
\rho_A(\br',t)
\rho_B(\br'',t)
    }_\text{NPT}.
\end{equation}
These density correlation functions from the NPT ensemble encodes the particle fluctuations inside the $\mu$VT ensemble of volume $V$, because
\begin{equation}
    \int\limits_V 
    d \br
    \rho_A^{(1)}(\br)
    = \avg{\na},
\end{equation}
\begin{equation}
    \int\limits_V d\br' 
    \int\limits_V d\br''
    \rho_{AB}^{(2)}(\br',\br'')
    = \avg{\na \nb},
\end{equation}
\begin{multline}
    \int\limits_V d\br' 
    \int\limits_V d\br''
    \left(
    \rho_{AB}^{(2)}(\br',\br'')
    -\rho_A^{(1)}(\br') \rho_B^{(1)}(\br'')
    \right)\\
    = \avg{\Delta\na \Delta\nb}.
    \label{eq:rhoAB}
\end{multline}

One can do a Fourier expansion of the instantaneous density field in space $\br$ inside the $\mu$VT ensemble with volume $V$, e.g.
\begin{equation}
\widetilde{\rho}_A(\bk,t) = 
\int_V d\br \rho_A(\br,t) \exp(i \bk \cdot \br)
= \sum_{i=1}^{N_A} \exp(i \bk \cdot \br_{i_A}(t)).
\label{eq:rhokt}
\end{equation}
As the density field is a real function, $\widetilde{\rho}_A(-\bk,t)=\widetilde{\rho}_A^\star(\bk,t)$, where the latter is the complex conjugate.

The static structure factor between two types of particles (A--A, A--B, or B--B) is defined as
\begin{equation}
        S_{AB}(\bk) = \dfrac{1}{\sqrt{\avg{\na} \avg{\nb}}}\avg{\widetilde{\rho}_A(\bk,t)\widetilde{\rho}_B(-\bk,t)}.
\end{equation}
It is easy to verify that the $S_{AB}(\bk)$ is the Fourier expansion of $\rho^{(2)}_{AB}(\br',\br'')$, i.e.
\begin{multline}
    S_{AB}(\bk) = 
    \dfrac{1}{\sqrt{\avg{\na} \avg{\nb}}}\\
    \int_V d\br'  \exp(i \bk \cdot \br')
    \int_V d\br''  \exp(i \bk \cdot \br'')
    \avg{\rho_A(\br',t)\rho_B(\br'',t)}.
    \label{eq:ft-sab}
\end{multline}

Combining Eqn.~\eqref{eq:rhoAB} with ~\eqref{eq:ft-sab}, 
the structure factor is related to the particle number fluctuations via
\begin{equation}
S_{AB}^0 \equiv
    \lim_{\bk \rightarrow 0} S_{AB}(\bk) =
    \dfrac{\avg{
    \Delta \na \Delta \nb}}
    {\sqrt{\avg{\na}\avg{\nb}}}.
    \label{eq:S-fluc}
\end{equation}
Furthermore, the Kirkwood--Buff integral (KBI)~\cite{Kirkwood1951} between components 
$A$ and $B$ is
and related to the structure factor by
\begin{equation}
    G_{AB} = \dfrac{1}{\sqrt{\ca\cb}} 
    \left(S_{AB}^0-\delta_{AB}\right).
\end{equation}

Plugging Eqn.~\eqref{eq:S-fluc} into Eqn.~\eqref{eq:dmudc}, one obtains
\begin{equation}
    \left(\frac {\partial \ma}{\partial c_A}\right)_{T,P}=
        \dfrac{k_BT}{\ca}
    \left[
    \dfrac{1}
    {  S_{AA}^0-  S_{AB}^0\sqrt{\ca/\cb}
    }
    \right].
\end{equation}

Importantly,
although the Fourier expansions above are performed inside the volume $V$ of the $\mu$VT ensemble,
for isotropic liquid with translational invariance,
$S_{AB}(\bk)$ should be the same in any parts of the larger NPT ensemble--that includes the whole volume of the NPT ensemble.
To obtain $S_{AB}(\bk)$ from NPT simulations of finite fluid systems with 
periodic boundary conditions, 
one can only let one dimension of the simulation box fluctuate with barostat, while only collecting $S(\bk)$ for $\bk$ along the plane of the two fixed dimensions.
Even better, one can also perform the Fourier expansion on the scaled coordinates, and obtain the static structure factors using
\begin{equation}
    S_{AB}(\bk) = \dfrac{1}{\sqrt{N_A N_B}}
    \avg{\sum_{i=1}^{N_A} \exp(i \bk \cdot \hat{\br}_{i_A}(t))
    \sum_{i=1}^{N_B} \exp(-i \bk \cdot \hat{\br}_{i_B}(t))}
\end{equation}
where $\hat{\br}(t) = \br(t)\avg{l}_\text{NPT}/l(t)$ and $l(t)$ is the instantaneous dimension of the supercell.
The scaling procedure is rigorous at the thermodynamic limit, where the NPT and the NVT ensembles are equivalent.

To determine $\lim_{\bk \rightarrow 0} S_{AB}(\bk)$
from MD simulations of finite system sizes, one can compute $S(\bk)$ at small $\bk$ under the NPT ensemble, and extrapolate to the $\bk \rightarrow 0$ case using the Ornstein--Zernike form~\cite{barrat2003basic}:
\begin{equation}
    S_{AB}(\bk) = 
    \dfrac{S_{AB}^0}{1+k^2\xi_A\xi_B}.
    \label{eq:oz-2b}
\end{equation}

Finally, to compute chemical potentials, one can run multiple equilibrium NPT simulations with different concentrations and then obtain $\ma(\ca=\avg{c_A}_\text{NPT})$ using numerical integration with respect to $\ln\ca$:
\begin{multline}
    \ma(c_A) = \ma^0 + \kb T \ln(\dfrac{c_A}{c_A^0}) \\
    + \kb T \int_{\ln c_A^0 }^{\ln c_A} d \ln(c_A)
        \left[
    \dfrac{1}
    {  S_{AA}^0-  S_{AB}^0\sqrt{\ca/\cb}
    }-1
    \right],
    \label{eq:S-integral}
\end{multline}
referenced to the chemical potential $\ma^0$ at a standard molar concentration $c_A^0 $. 
One can conveniently select this reference to be the pure state.
Strictly speaking, Eqn.~\eqref{eq:S-integral} provides the relative chemical potential instead of the absolute value, but only the former is a physical observable.
The first two terms on the right hand side of Eqn.~\eqref{eq:S-integral} is the ideal-mixture chemical potential $\mu^{id}$, 
and the third term is the excess chemical potential $\mu^{ex}=\kb T \ln (\gamma_A)$ where $\gamma_A$ is the activity coefficient of the solute A.
The ratio $\gamma'_A = \dfrac{1}
{  S_{AA}^0-  S_{AB}^0\sqrt{\ca/\cb}
    }$ is related to the activity coefficient by 
    $\gamma'_A = 1+d\ln (\gamma_A)/d\ln \ca$.

In addition, although one can evaluate $\ma(c_A)$ and $\mb(c_B)$ separately using Eqn.~\ref{eq:S-integral} from the same simulations, 
one can obtain $\ma$ from $\mb$ employing the Gibbs-Duhem equation under constant P,T conditions~\cite{mouvcka2013molecular}:
\begin{equation}
    N_A d\ma + N_B d\mb =0.
    \label{eq:GD}
\end{equation}
To automatically satisfy the Gibbs-Duhem with the S0 method, one can use
\begin{equation}
    \left(\frac {\partial \ma}{\partial \ln(\chi_A)}\right)_{T,P}=
    \dfrac{k_BT}
    {\chi_B  S_{AA}^0
    +\chi_A  S_{BB}^0
    - 2\sqrt{\chi_A \chi_B}S_{AB}^0
    },
    \label{eq:s-integral-gd}
\end{equation}
where $\chi_A$ and $\chi_B = 1-\chi_A$ are the molar fraction of $A$ and $B$.

It is worthwhile discussing the finite size effects in the current approach, and in particular, 
the difference with the KBI method.
Typically, when using KBI,
one either starts from the pair correlation functions,
or one collects $\na$ and $\nb$ inside a fixed volume $V$ during NPT simulations~\cite{CortesHuerto2016,Brten2021}.
However, the open boundary of $V$ imposes very large finite size effects~\cite{CortesHuerto2016},
and even for $\mu$VT systems with hundreds of thousands of atoms,
a significant fraction are lying on the boundary~\cite{Heidari2018}.
Without finite size corrections, the KBI approach is hardly applicable~\cite{CortesHuerto2016,Brten2021}.
In contrast, computing $S(\bk)$ in an NPT ensemble avoids such boundary effects.
In addition, the effects coming from using a finite wavelength ($2\pi/k$) is partly corrected by the physically-inspired extrapolation (Eqn.~\eqref{eq:oz-2b}).

\paragraph{Binary Lennard-Jones system}
A binary mixture
(A, B) of Lennard-Jones (LJ) fluids was simulated using a purely
repulsive 6-12 LJ potential truncated and shifted with
cutoff radius $2^{1/6}\sigma$. 
The potential parameters are  $\sigma_{AA} = \sigma_{BB}$ = $\sigma_{AB} = 1$, and $\epsilon_{AA} = 1.2$, $\epsilon_{BB} = 1.0$,
with $\epsilon_{AB} = (\epsilon_{AA} + \epsilon_{BB})/2)$. 
Constant temperature and pressure was enforced through a stochastic velocity rescaling
thermostat
and Nose-Hoover barostat.
Three systems
sizes, 4000, 23328 and 108000 total number of atoms, were considered in the range
of mole fractions of A particles, $\chi_A = 0.05, \ldots ,1.0$. 
Simulations were carried
out using LAMMPS~\cite{Plimpton1995} with a time step of $10^{-3}$.
Total simulation time was $10^6$ simulation steps 
for the two larger systems, and $10^7$ for $N=4000$.
To compute $S(\bk)$, we
collected a snapshot per 2000 steps of the trajectory.
\begin{figure}[tbp]
    \centering
    
    \begin{tabular}{@{}p{0.45\textwidth}@{\quad}p{0.45\textwidth}@{}}
    \subfigimg[width=\linewidth]{\figLabel{a}}{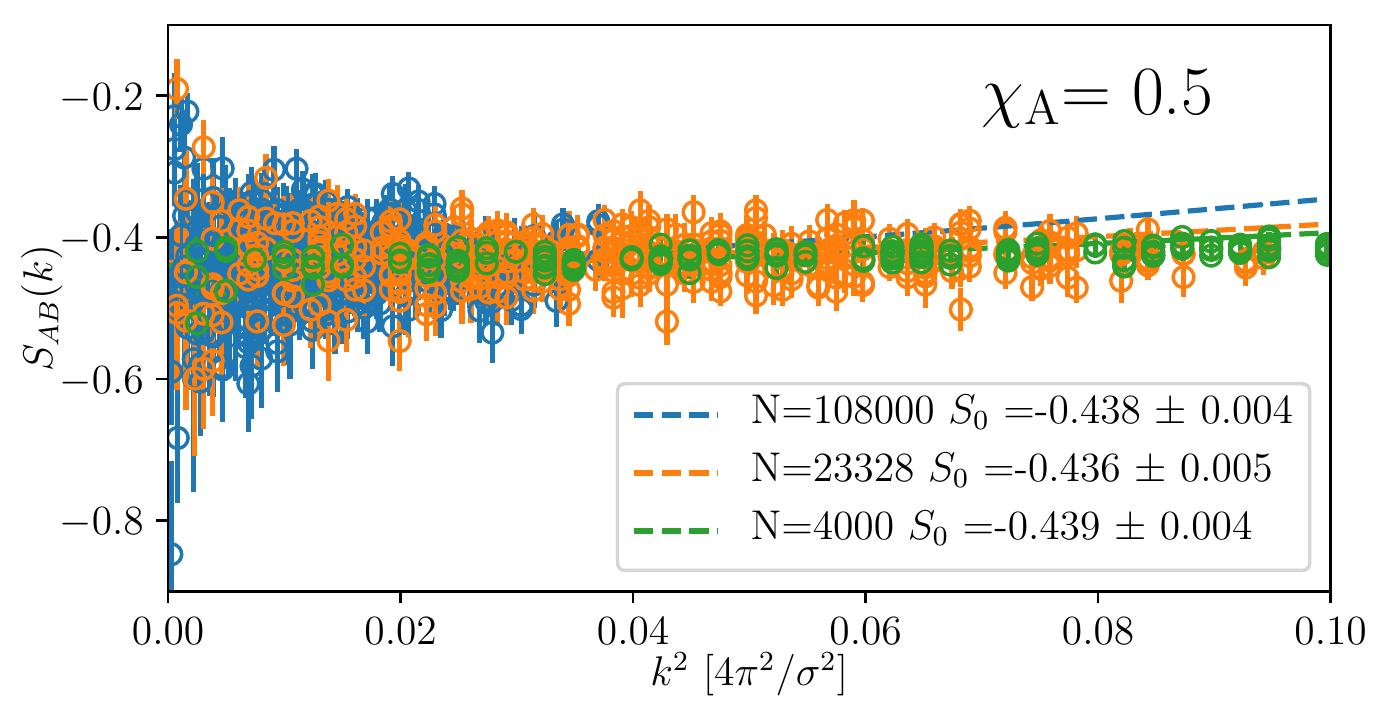} \\
    \subfigimg[width=\linewidth]{\figLabel{b}}{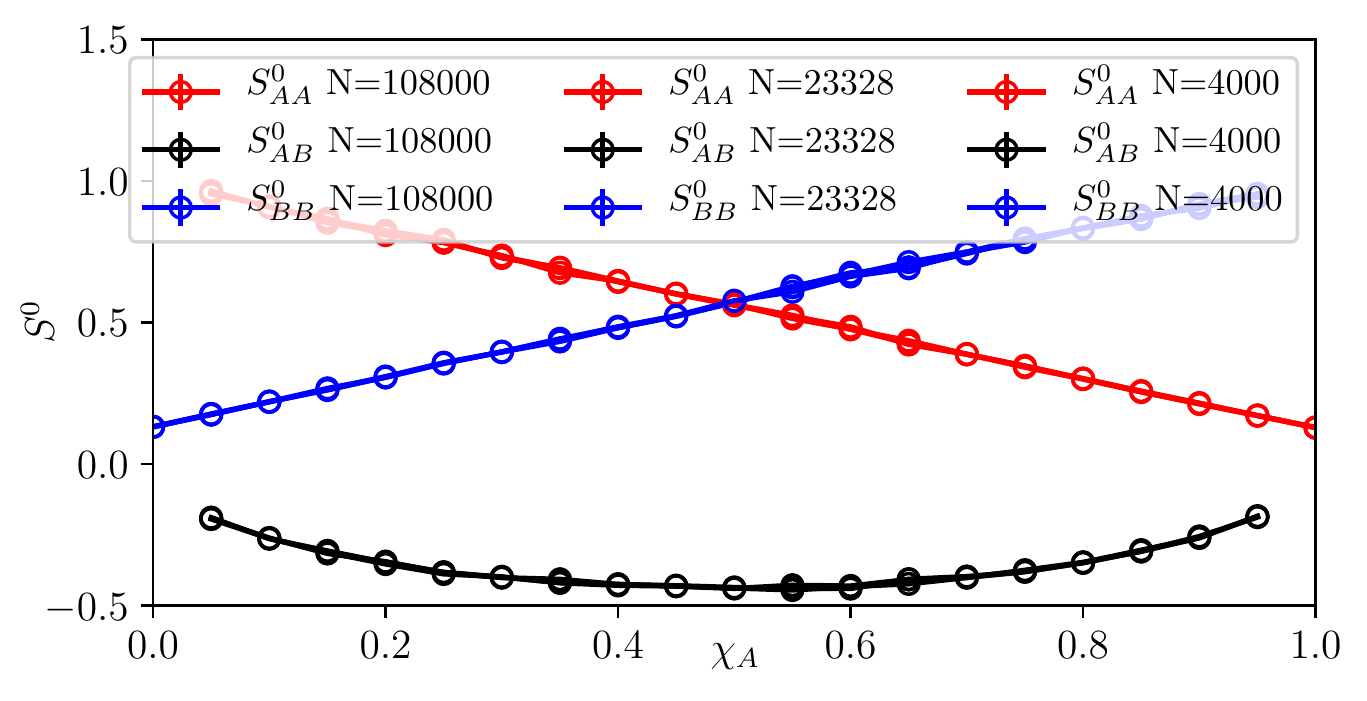}  \\
    \subfigimg[width=\linewidth]{\figLabel{c}}{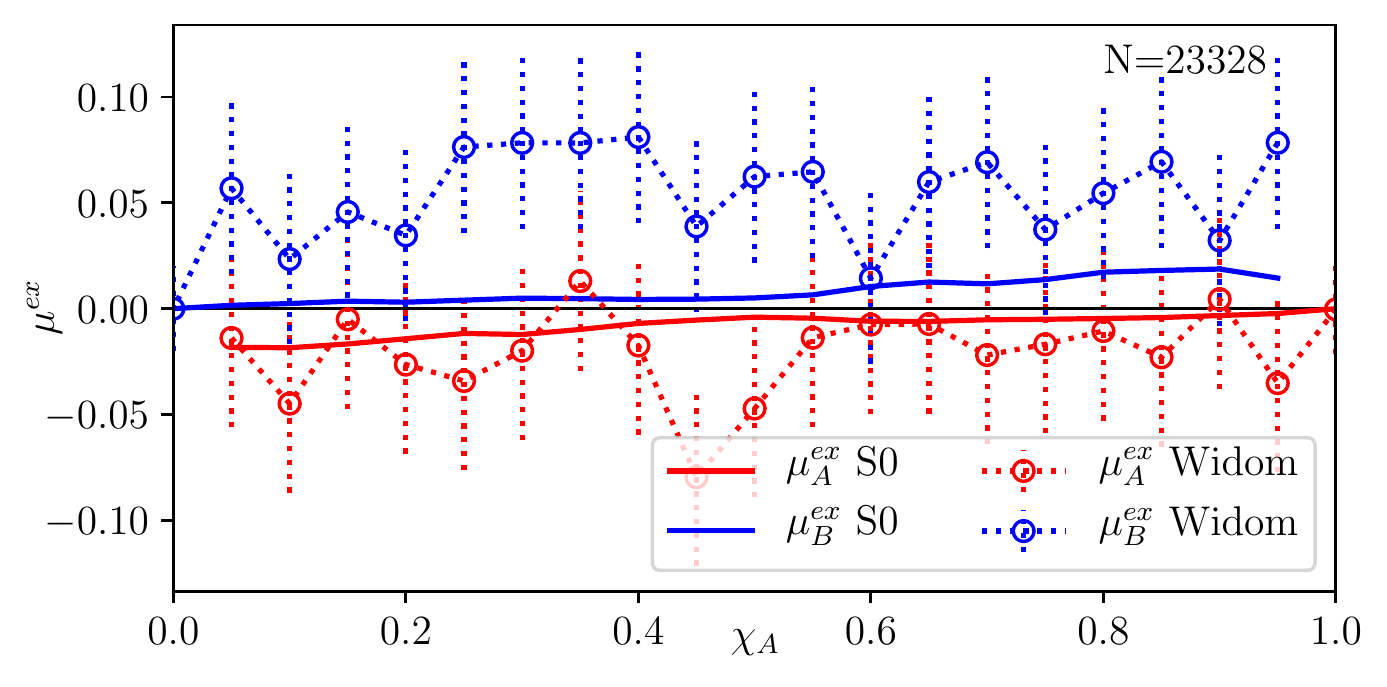}
  \end{tabular}
    \caption{
    Analysis the chemical potentials of the binary LJ system using the S0 method.
    \figLabelCapt{a} $S_{AB}(\bk)$ computed from NPT simulations with $T=1.2$ and $P=2$, using different system sizes $N$.
    The dashed curves are the corresponding fits using Eqn.~\eqref{eq:oz-2b}.
    \figLabelCapt{b} $S_{AA}^0$, $S_{AB}^0$ and $S_{BB}^0$ at different molar fractions of $A$, computed using three system sizes.
    \figLabelCapt{c} The excess chemical potentials of particle A and B at different $\chi_A$, computed using the S0 method and the Widom particle insertion.}
\label{fig:lj}
\end{figure}

\figrefsub{fig:lj}{a} shows $S_{AB}(\bk)$ as a function of $k^2=\abs{\bk}^2$ computed from NPT simulations at different system sizes for a selected molar fraction of particles.
The dashed curves are fits to Eqn.~\eqref{eq:oz-2b} with a maximum cutoff in the wavevector $k_\text{cut}=2\pi \times 0.2\sigma^{-1}$,
although we found the fitted values for $S^0$ are insensitive to this choice.
Even with a small size of 4000 particles the estimate for $S_{AB}^0$ is converged.
Such insensitivity to system size is again confirmed from \figrefsub{fig:lj}{b}, which shows the
the extrapolated values of $S_{AB}^0$ at different molar fractions. 
As a benchmark, in \figrefsub{fig:lj}{c} we compare the excess chemical potential $\mu^{ex}_A$ and $\mu^{ex}_B$, with the reference $\ca^0$ and $\cb^0$ set to the concentrations of pure A and pure B (Eqn.~\eqref{eq:S-integral}),
computed using the S0 method and Widom particle insertion~\cite{frenkel2001understanding}.
Widom insertion simulations were performed the same system size (23328 atoms),
$4\times10^6$ time step and one particle insertion each time step.
Both methods captures the fact that $\mu^{ex}$ for both A and B are larger when $\chi_{A}$ is higher, because A particles have a stronger repulsive core,
but the S0 method has much better statistical efficiency.

\begin{figure}
\includegraphics[width=0.45\textwidth]{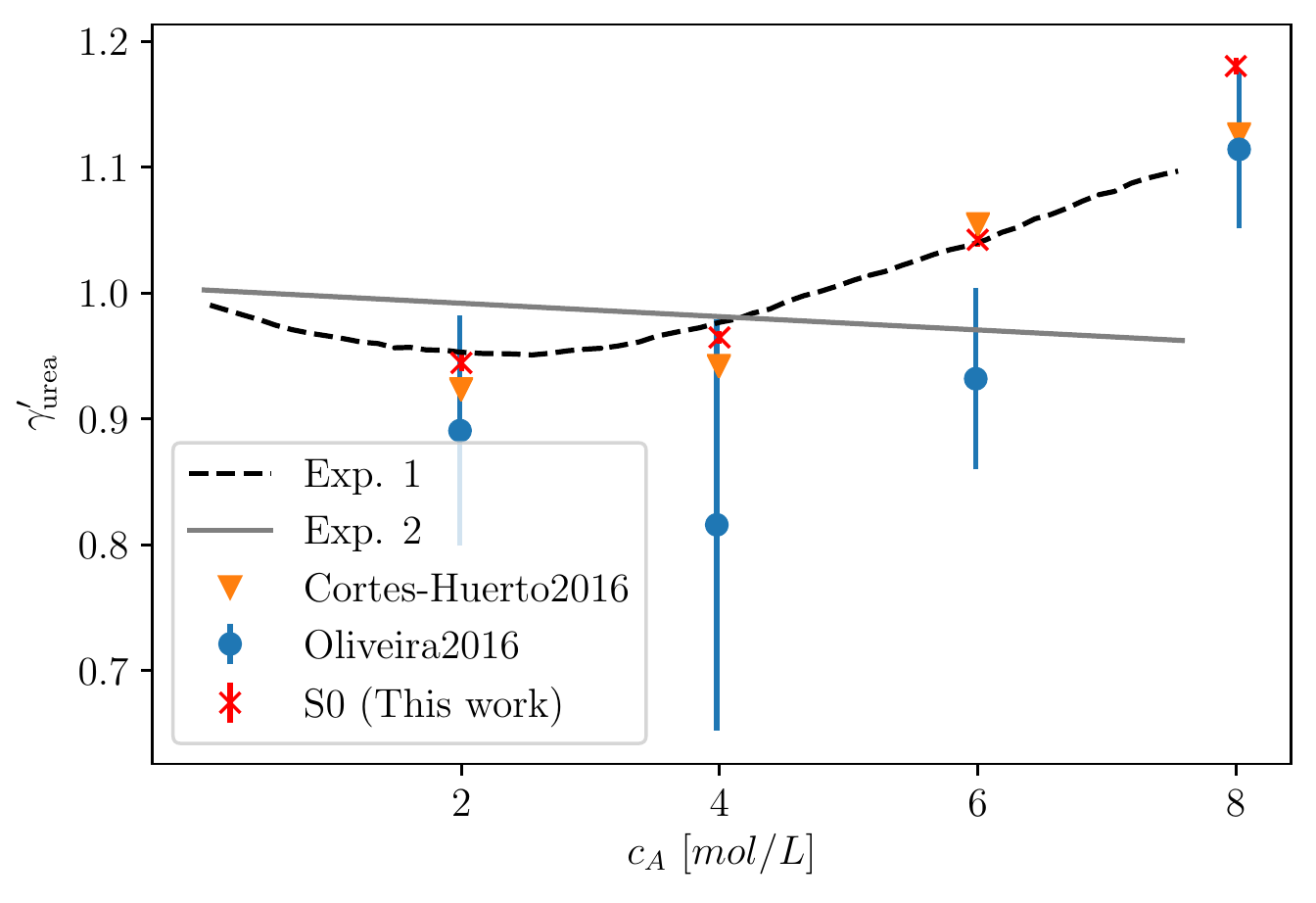}
\caption{
Comparison of the derivative of the activity coefficient ($\gamma'_\text{urea}$) with the previous simulation results~\cite{deOliveira2016,CortesHuerto2016} and experiments (Exp.1~\cite{Stokes1967,Weerasinghe2003}, Exp.2~\cite{Miyawaki1997}).
}
\label{fig:urea}
\end{figure}

\paragraph{Urea in water}
We analyzed the MD trajectories of urea/water mixtures from Ref.~\cite{deOliveira2016}, simulated 
using the Kirkwood-Buff derived force field~\cite{Weerasinghe2003} and SPC/E water at 1 atm pressure and 300 K temperature.
Four urea molar concentrations were considered (2, 4, 6, and 8 M),
and the system size is 13000-16000 total number of urea and water molecules.
The comparison of the derivative of the activity coefficient $\gamma'_\text{urea} = 1+d\ln (\gamma_\text{urea})/d\ln c_\text{urea}$ is in Fig.~\ref{fig:urea}.
The S0 results closely agree with the Ref.~\cite{CortesHuerto2016} that uses KBI with finite size corrections, and with Exp.~1~\cite{Stokes1967,Weerasinghe2003}. 

\paragraph{NaCl aqueous solution}
Simulations of NaCl water solutions
at different molar concentrations (0.1-9.3M) were performed using LAMMPS~\cite{Plimpton1995}
at 298.15~K and 1~bar.
The JC/SPC/E~\cite{joung2008determination} forcefield was used, with the Lennard-Jones
interactions truncated at 1 nm,
long-range
Coulomb interactions treated using a particle-particle particle-mesh solver,
and the constraints for the rigid water molecules enforced by SHAKE.
A fix amount of 32640 water molecules together with 128-10880 NaCl ion pairs are used.
The timestep is 2~fs, the total steps are 2000000, and the
Nose-Hoover thermostat and barostat are used.

\begin{figure}
\includegraphics[width=0.45\textwidth]{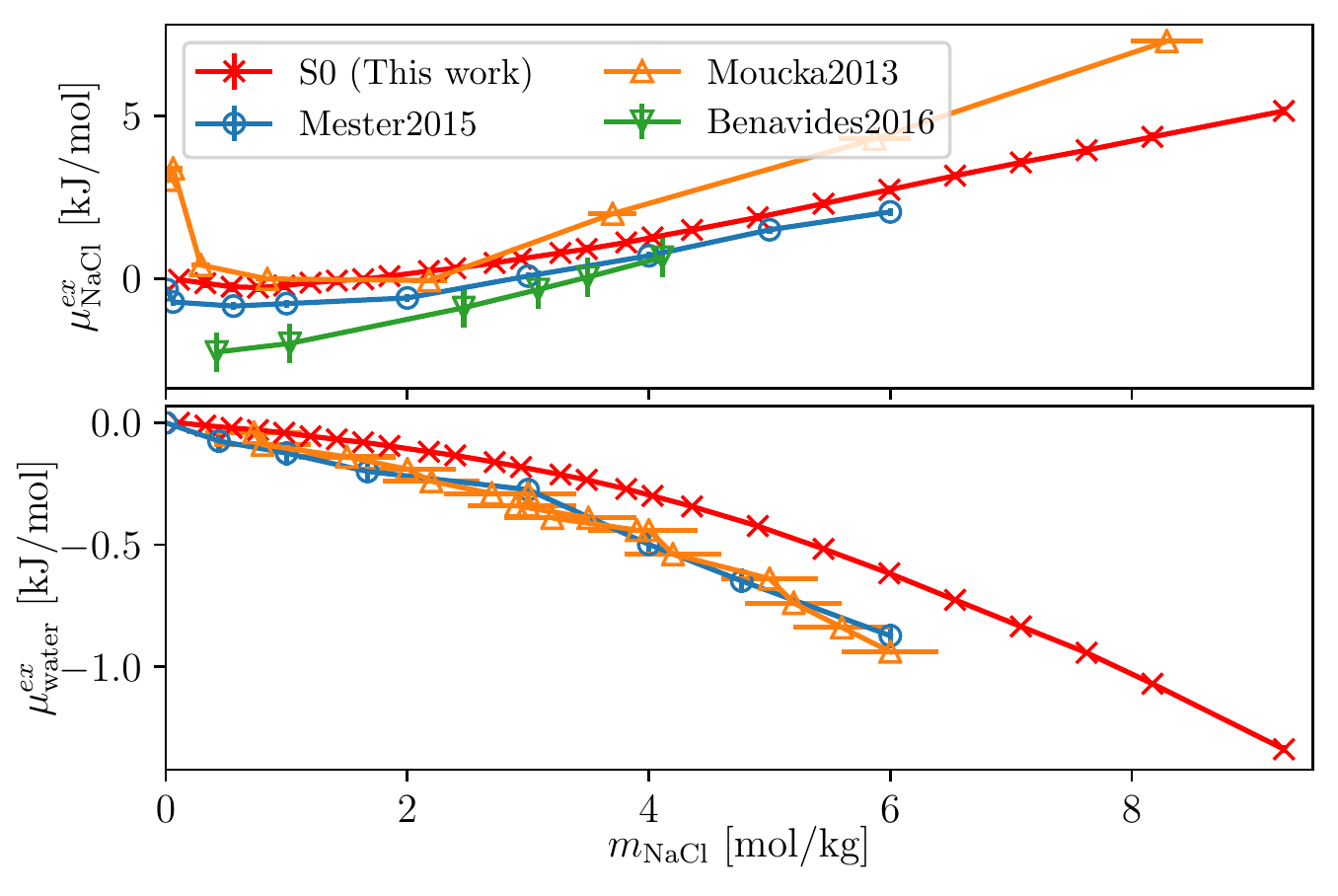}
\caption{
The excess chemical potentials of NaCl ion pairs (upper panel) and water molecules (lower panel) as a function of molality,
compared with previous simulation results~\cite{moucka2011molecular,Mester2015,benavides2016consensus} all employing the JC/SPC/E~\cite{joung2008determination} forcefield.
}
\label{fig:nacl}
\end{figure}

Fig.~\ref{fig:nacl} shows the excess chemical potentials for NaCl ion pairs ($\mu^{ex}_\text{NaCl}$) and water ($\mu^{ex}_\text{water}$) at different salt molality $m$ (mol NaCl/kg water)
calculated using the S0 method,
which was integrated using Eqn.~\eqref{eq:S-integral}, and we found that employing Eqn.~\eqref{eq:s-integral-gd} rendered fully consistent values.
$\mu^{ex}_\text{NaCl}$ and $\mu^{ex}_\text{water}$ are compared with previous results computed
using
Osmotic Ensemble Monte Carlo~\cite{mouvcka2013molecular}, 
the Bennett acceptance ratio method~\cite{Mester2015}, and thermodynamic integration~\cite{benavides2016consensus}.
Note that the values of $\mu^{ex}_\text{NaCl}$ depends on the standard chemical potential reference,
and each study handles this differently (we set $\mu^{ex}_\text{NaCl}(m=0.1)=0$),
so only the change as a function of $m$ is physically meaningful. 
All four studies compared in Fig.~\ref{fig:nacl} are fairly consistent, while the S0 method agree particularly well with Ref.~\cite{Mester2015}.

\paragraph{High-pressure carbon-hydrogen mixture}
The implications for this C-H mixture in the context of planetary science will be discussed in Ref.~\cite{cheng2022ch},
and here we focus on how adding hydrogen to the liquid carbon changes the chemical potential of carbon atoms.
We used a machine learning potential developed in Ref.~\cite{cheng2022ch},
and performed NPT simulations at different $N_\text{C}/N_\text{H}$ ratios at T=5000~K and P=100~GPa.
The system sizes are between 11232 and 82944 total number of atoms.

\begin{figure}
\includegraphics[width=0.45\textwidth]{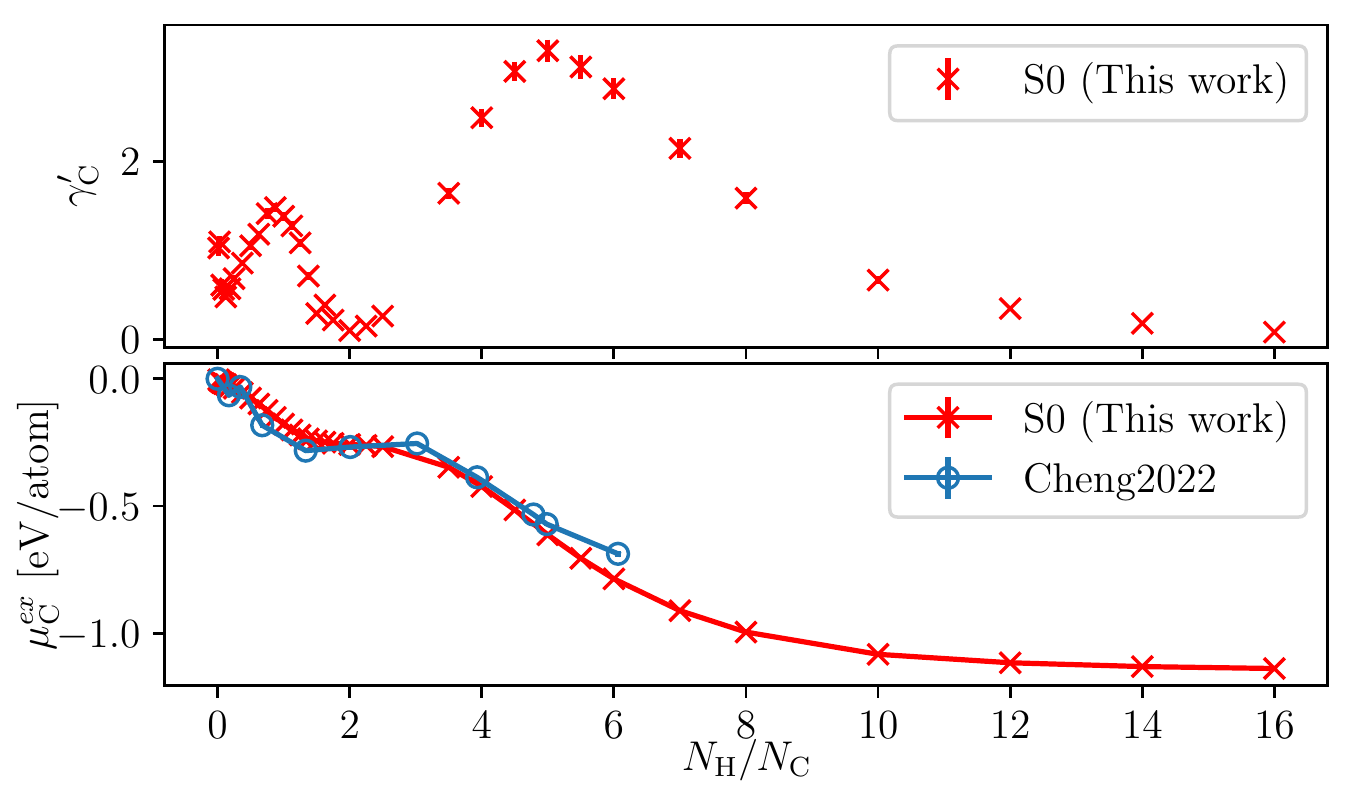}
\caption{
The derivative of the activity coefficient $\gamma'_\text{C}$ (upper panel),
and the excess chemical potential $\mu_\text{C}^{ex}$ (lower panel)
at different molar fraction of C, at T=5000~K and P=100~GPa.
$\mu_\text{C}^{ex}$ are referenced to the bulk liquid carbon,
and compared with values from coexistence simulations~\cite{cheng2022ch}.
}
\label{fig:CH}
\end{figure}

In Fig.~\ref{fig:CH} we show the derivative of the activity coefficient ($\gamma'_\text{C}$) and excess chemical potential $\mu^{ex}_C$ of carbon at different molar fraction.
The S0 method predicts $\mu^{ex}_C$ that are in good agreement with the results obtained by coexistence simulations~\cite{cheng2022ch},
but it can be used for the regime of low carbon concentration which becomes prohibitive using the coexistence approach.
Interestingly, $\gamma'_\text{C}$ shows nonmonotonic behavior with a minimum at $N_\text{H}/N_\text{C}=2$, and maxima at $N_\text{H}/N_\text{C}=1$ and 4.
These magic numbers are probably related to the chemical bonds between C and H.

To conclude, we present the S0 method to compute chemical potentials of mixtures just from equilibrium MD simulations at the NPT ensemble,
by simply computing static structure factors.
We demonstrate the generality and robustness of the S0 method on diverse systems including a model LJ system,
organic molecule in water,
aqueous electrolyte, and a high-pressure solution.
In principle, the S0 method is also applicable to larger molecules such as polymers, or large-molecule solvents.
Compared with the previous methods such as particle insertion~\cite{allen2012computer,smit1989calculation}, thermodynamic integration~\cite{sanz2007solubility,paluch2010method}, the S0 method is more generally applicable and  works particularly well for dense fluids with complex interactions and high solute concentrations.
We envisage
the S0 method will largely simplify the computation of the chemical potentials of complex solutions,
and make them routine endeavors.

\textbf{Acknowledgements}
I thank Daan Frenkel for providing feadback on an early draft and for stimulating discussions.
I thank Debashish Mukherji and Robinson Cortes-Huerto  for kindly sharing the trajectories
for urea water mixtures.
I thank Aleks Reinhardt for useful suggestions on the manuscript.

\textbf{Data availability statement}
All PYTHON scripts and simulation input files generated for the study are in the
SI repository (url to be inserted upon acceptance of the paper).

\end{document}